\documentclass[12pt,a4papper]{article}
\usepackage{graphics}
\usepackage{graphicx}
\usepackage{amssymb}
\usepackage{color}

\renewcommand{\r}{{\bf r}}
\newcommand{\astr}{_{\rm astr}}
\newcommand{\atom}{_{\rm atom}}
\newcommand{\dif}{{\rm d}}

\begin{document}

\title{On the compatibility of a proposed explanation of the Pioneer anomaly with the cartography of the solar system}
\author{Antonio F. Ra\~nada\thanks{Facultad de
F\'{\i}sica, Universidad Complutense, 28040 Madrid, Spain. E-mail
afr@fis.ucm.es, corresponding author.} \and Alfredo
Tiemblo\thanks{Instituto de Matem\'aticas y F\'{\i}sica
Fundamental, Consejo Superior de Investigaciones Cient\'ificas,
Serrano 113b, 28006 Madrid, Spain.}}
\date{15 August 2009}
\maketitle

\tableofcontents

\newpage

\begin{abstract}

We analyze here the reasons why an explanation of the Pioneer
anomaly proposed by the authors is fully compatible with the
cartography of the solar system. First, this proposal posits that
the phenomenon is an apparent acceleration, not a real one, caused
by a progressive desynchronization of the astronomical and the
atomic clock-times, after they had been synchronized at a previous
instant. The desynchronization could be caused by a coupling between
the background gravitation and the quantum vacuum. Therefore, the
standard argument for the incompatibility of the Pioneer
acceleration and the values of the planets' orbits radii cannot be
applied. Second, this proposal gives exactly the same results for
radar ranging observations as standard physics. Hence, it cannot be
in conflict with the very precise cartography of the solar system
determined by NASA's Viking mission. Otherwise stated, while this
proposal predicts apparent changes in the velocities of the
spaceships and in the frequencies of Doppler observations, as really
observed, it does not affect the values of the distances in any way
whatsoever. Note that an acceleration between the astronomical and
the atomic clock-times ({\it i. e.} a progressive desynchronization)
can not be discarded a priori as long as we will lack a theory of
quantum gravity.

\end{abstract}

\section{Introduction: the Pioneer anomaly}
The purpose of this note is to analyze the reasons why
 an explanation of the Pioneer anomaly proposed by the
authors is fully  compatible with the cartography of the solar
system \cite{RT08,RT09}. In fact it gives exactly the same values as
NASA's Viking mission for the distances from any planet to the
others and to the sun. This is important since the failure to find a
model with this compatibility has been a major difficulty to explain
this intriguing phenomenon,  reported in 1998 by Anderson {\it et
al.}, although they had been studying it since 1980
\cite{And98,And02}. It consists in an
 adiabatic frequency blue drift of the two-way radio signals
from the Pioneer 10 and 11 (launched in 1972 and 1973), manifest
in a residual Doppler shift that increases linearly with time as
\begin{equation} \dot{\nu}/\nu =2a_{\rm t}\,,\quad \mbox{or}\quad
\nu =\nu_0\,[1+2a_{\rm t}(t-t_0)]\,,\label{10}\end{equation} where
$t_0$ is an initial time and $2a_{\rm t}= (5.82\pm 0.88)\times
10^{-18}\mbox{ s}^{-1}\simeq(2.53\pm 0.38) H_0$, $H_0$ being the
Hubble constant (overdot means time derivative). It must be stressed
that the signal found by Anderson {\it et al.} is very clean and
well-defined. Because of its linearity in time, the residual
frequency $\nu$ of the ship appears as a straight line in a plot
($\nu,\,t$) \cite{And98}. This strongly suggests that the anomaly is
the signature of a new phenomenon, probably with a cosmological
origin, so that the curve $\nu(t)$ can be accurately approximated by
its tangent for a scale of just a few decades.

The simplest interpretation of eqs. (\ref{10}) is that the ship was
submitted to a constant force directed towards the sun. However this
would be in conflict with the cartography of the solar system and
with the equivalence principle. In view of this difficulty, the
discoverers held the view that ``the most likely cause of the effect
is an unknown systematics". This systematics was not found, however,
in spite of several different analysis of the data. More recently,
some researchers put their hopes in  the so called thermal model,
which assumes that the effect is due to non isotropic radiation by
the spaceship. But it is difficult to imagine that such radiation
could give a simple and neat straight line in the diagram
($\nu,\,t$). In fact, the phenomenon is still unexplained many years
after its discovery.

\section{Summary of our proposal}

We showed in reference \cite{RT09} that the analysis of a coupling
between the quantum vacuum and the background gravitation presented
in \cite{RT08} gives a solution to the anomaly, which is free of
internal contradictions and does not conflict with any established
physical law or principle. The reader is referred to \cite{RT08} for
the details.  In this section, we give just a terse summary of our
proposal as follows: i) a coupling between the quantum vacuum and
the background gravity that pervades the universe is unavoidable
because of the long range and universality of the gravity; ii) the
fourth Heisenberg relation implies then that this coupling must
cause a progressive desynchronization of the astronomical and the
atomic clocks, after they had been synchronized at any arbitrary
previous instant, in such a way that the former decelerate
adiabatically with respect to the later; iii) since gravitational
theories use astronomical time, say $t\astr$, and observers use
atomic time, say $t\atom$ (they are using devices based on quantum
physics), this desynchronization necessarily causes a discrepancy
between
 theory and observation. The consequence is that the observed velocity of the
 spaceship is smaller than the predicted one,
 in such a way that the Pioneer
 seems to lag behind its expected position. In our proposal,
 therefore, the  anomaly is a cosmological effect.

 We must underscore that the possibility that
the two times are accelerating with respect to one another  cannot
be discarded a priori, as long a we will lack a unified theory of
gravitation and quantum physics. Also that although the best
election for $t\astr$ is probably the ephemeris time with
relativistic corrections,  any other time based on the motion of
celestial bodies could be used instead.

Our model accepts the following phenomenological hypothesis: the
empty space can be considered as a substratum, a transparent optical
medium, characterized by a permittivity and a permeability due to
the sea of virtual pairs that are continuously created and
destroyed. As a consequence, therefore, of the coupling between the
quantum vacuum and the background gravity, the existence must be
admitted of some kind of adiabatic progressive modification of the
structure of the quantum vacuum in the expanding universe. Let $\Psi
(t)$ be the background gravitational potential that pervades the
universe, in the approximation that all the mass-energy is uniformly
distributed. Assuming that $\Psi=0$, it follows  from the fourth
Heisenberg relation that the average lifetime of a pair with energy
$E$ can be taken to be $\tau _0=\hbar /E$. On the other hand, if
$\Psi\neq 0$,  the pairs acquire an extra energy $E\Psi$ so that
their lifetime and number density must depend on the potential as
\begin{equation}
\tau _\Psi=\hbar /(E+E\Psi)=\tau _0/(1+\Psi);\quad {\cal N}_\Psi
={\cal N}_0/(1+\Psi)\label{10a}\end{equation} (see \cite{RT08},
section 4.) As is seen, the fourth Heisenberg relation implies
that the gravitational potential affects the density of the
quantum vacuum in such a way that the more negative is the
potential, the larger is the number density of pairs. As a
consequence the optical properties of empty space must depend on
the potential, {\it i. e.} on time, including  its permittivity
and permeability. As shown in \cite{RT08}, it follows that the
astronomical time $t\astr$ and the atomic time $t\atom$, which are
equal in the absence of gravity, accelerate with respect to one
another, so that
\begin{equation}\dif t\atom /\dif
t\astr=u(t\astr)\neq 0,\label{20a}\end{equation} $u$ being a
function of time which we call the ``march" of $t\atom$ with
respect to $t\astr$. This variation must be very small, otherwise
it would have been detected before, so that $u$ is very close to
one. At first order in the variation of time, this march can be
expressed as
\begin{equation} u=1+a(t\astr -t_{\rm
astr,\,0}),\quad a={\dif ^2t\atom \over \dif t\astr
^2},\label{30a}\end{equation} where $a$ is a positive inverse time
depending on the potential that corresponds to $2a_{\rm t}$ in
\cite{And98} and $t_{\rm astr,\,0}$ is an initial insant at which
the two times are synchronized. Since $a$ must be very small, the
phenomenon is adiabatic. Note that, at first order, it is not
necessary to specify the kind of time in the RHS of the first
equation (\ref{30a}). As a consequence, the speed of light is
constant if defined or measured with atomic time but increases
with $t\astr$. In fact, since $v\astr =\dif \ell/\dif t\astr$ and
$v\atom =\dif \ell /\dif t\atom$,
\begin{equation} v\astr
=uv\atom; \quad c\astr =uc.\label{40a}\end{equation} Because $u>0$
for $t>t_0$, it happens that $v\atom <v\astr$ for $t>t_0$, so that
the atomic velocity is smaller than the astronomical one, this
explaining why the ship seems to lag behind its expected position.

A comment is suitable here: Our proposal might seem strange at
first sight but note that the effect remains unexplained more than
25 years after its discovery, so that it would seem stranger yet
that, once its solution is found, it didn't {\it seem} strange.

\section{The problem of the compatibility with the cartography of
the solar system} From the very beginning it was clear that the
phenomenon poses some difficult problems. First of all and since a
similar effect was not observed in the planets, it seemed that it
could affect a small body as the Pioneer but had no action on more
massive ones, what is incompatible with the equivalence principle,
a cornerstone of gravitation theories. Second, it was thought that
the Pioneer acceleration $a_{\rm P}$ should be due to an
attractive force towards the Sun, so that the radii of the
planets's orbits should be smaller than their well known values.
Third, any  explanation of the effect should face a difficult
proof: to agree with the extremely precise determination of the
solar system cartography achieved by Viking's mission.

Let $a_{\rm P}$ be a real acceleration. In that case, the radial
equation of a planet should be
\begin{equation} \ddot{r}=-GM/r^2+J^2/r^3-a_{\rm
P},\label{30}\end{equation} with standard notation \cite{Ran04}. In
the limit case of circular orbits the RHS vanishes so the radius
changes to $r+\Delta r$ with
\begin{equation} \Delta r=-ra_{\rm P}/a_{\rm N},\label{40}
\end{equation}
at first order in $a_{\rm P}$, where $a_{\rm N}$ is the Newtonian
acceleration. For non circular orbits, one has $\Delta r=-J^6a_{\rm
P}/(GM)^4$. This is about $-21$ km and $-76$ km for Earth and Mars,
respectively, so that the radii of the orbits of these two planets
should be 21 km and 76 km smaller than their accepted values.
However, this effect is not observed. More precisely, Anderson {\it
et al.} conclude than any unmodelled radial acceleration acting on
Earth and Mars and larger than $0.1\times 10^{-10}\mbox{ m/s}^2$ is
unacceptable. Indeed, it would be in conflict with NASA's Viking
data which determine the difference and the sum of the Earth and
Mars orbital radii to about 100 m and 150 m accuracy, respectively,
as the discoverers of the anomaly underline in their first paper
\cite{And98}. Note, however, that {\it this irreproachable argument
assumes as a necessary condition that the Pioneer effect is an
anomalous but real acceleration}, {\it i. e.} an anomalous but real
force ($\Delta r$ is calculated by adding $a_{\rm P}$ to the
planets' radial equation of motion). Otherwise, the phenomenon would
not include necessarily a negative correction to the radii of the
orbits. In any case, this caused somehow a misunderstanding that
probably has been hindering the solution of the riddle: that it
follows plainly from the Pioneer data that the effect should
necessarily imply a reduction of the planets' radii. This was
loosely interpreted sometimes as an indication that no mechanical
model can explain the anomaly and fostered the search for non
mechanical alternatives.

\section{Why our proposal is fully compatible with the cartography of the solar
system}

The previous argument on the decrease of the radii of the planets'
orbits {\it cannot be applied to our work because what we affirm is
that the Pioneer anomaly has the same observational fingerprint, and
could well be the same thing, as the effect of the desynchronization
of the astronomical and the atomic clock-times.} In other words: we
propose that {\it the Pioneer anomaly effect is an apparent, non
real, acceleration, not caused by any force. In our view, the
Pioneer did not suffer any extra acceleration but followed
faithfully the equations of standard gravitation theories. These
theories use astronomical clock-time and predict that the ship's
trajectory must be given by a certain function parameterized by time
$\r=\r(t_{\rm astr})$, which is the one followed by the Pioneer.
However, as the observers use devices based on quantum physics, they
use quantum clock-time $t_{\rm atom}$, so that what they observe is
the same trajectory, although parameterized by a different time and
given by a different function $\r'=\r'(t_{\atom})$. The two are
related as $\r'(t\atom)=\r(t\astr)$. If the observers are unaware of
the difference between the two times, they would interpret their
data as the fingerprint of an unmodelled anomalous acceleration,
{\it i. e.} they would see an anomaly.}

In fact, if the observers are unaware that there are two times which
accelerate with respect to one another and, consequently, use
 only one  variable $t$, there must be
necessarily an anomaly because of the difference between the
observed and the calculated distances from the ship to the sun
$\Delta \r =\r _{\rm obs}-\r_{\rm theor}=\r' (t)-\r(t)\neq 0$.  Then
as (i) $t\atom
>t\astr$, after the synchronization of $t\atom$ and $t\astr$ at
a previous time $t_0$, and (ii) both $|\r|$ and $|\r'|$ are
increasing functions of time since the ship is receding from the
sun, it happens that
\begin{equation} |\r'(t\atom)|=|\r(t\astr)|<|\r(t\atom)|.\label{60}\end{equation}
Since the  observers accept standard gravitation and use atomic
time, they expected to measure $|\r (t\atom)|$, {\it i. e.} the
theoretical prediction for the distance to the sun expressed in
terms of $t\atom$, but obtained instead $|\r'(t\atom)|.$ Otherwise
stated: the observed distance of the ship to the sun is always
shorter than the distance calculated with standard gravitation
theories, see \cite{RT09}, pp. 10-12 and figure 1. In our view the
anomaly boils down to that. There is no real acceleration, just an
apparent one. This argument shows also that there is no problem with
the equivalence principle in our model.

Moreover, it is easy to see that our model can have no conflict
with the very precise cartography of the solar system obtained by
NASA's Viking mission. In a radar ranging experiment, the
observers send a light ray from an initial point $P_1$ at time
$t_{\rm atom,\,1}$ and detect its arrival to $P_2$, which can be
the same one as $P_1$, at time $t_{\rm atom,\,2}$. The distance
traveled by the light ray is simply $d=c(t_{\rm atom,\,2}-t_{\rm
atom,\,1})=\int_{t_{\rm atom,\,1}}^{t_{\rm atom,\,2}} c\atom
\,\dif t\atom$. It turns out then that, in our model, the value of
this distance does not depend on which time is used. In fact,
changing the time variable and since $v\atom =v\astr /u$ and $\dif
t\atom =u\,\dif t\astr$ (eqs. (\ref{20a}) and (\ref{40a})),
\begin{equation} d=\int_{t_{\rm atom,\,1}}^{t_{\rm atom,\,2}} c\atom \,\dif t\atom=
\int_{t_{\rm astr,\,1}}^{t_{\rm astr,\,2}} c\astr \,\dif t\astr
.\label{70}\end{equation} {\it This shows that the predictions of
our model on the cartography of the solar system are exactly the
same as in standard physics, independently of which time is used.
The model is thus fully compatible with the results obtained by
the Viking mission because, although it predicts changes in the
velocities of the spaceships and the frequencies in Doppler
observations, it does not affect the distances in any way
whatsoever.}

To summarize, our proposal  (i) is free from internal
contradictions; (ii) is not in conflict with any established
physical law or principle, at least as long as we lack a working
unified theory of gravitation and quantum physics; and (iii)  does
not affect the accepted cartography of the solar system.
Consequently it does offer {\it a} solution of the Pioneer anomaly
and must be further investigated, therefore, as a candidate for {\it
the} solution.

\section{Acknowledgements} We are grateful to profs. A. I. G\'omez de Castro, J. Mart\'{\i}n
 and J. Us\'on for helpful discussions.

\end{document}